\documentclass[aps,twocolumn,amsmath,amssymb,pre,floatfix]{revtex4-1}
\usepackage{amssymb}
\usepackage{amsmath}
\usepackage{graphicx}
\usepackage[table]{xcolor}
\usepackage{epsfig}
\usepackage{dcolumn}
\usepackage{bm}
\usepackage{array}
\usepackage{multirow}
\begin{document}

\title{Second critical point of water in supercooled confined water in L,L-diphenylalanine micro/nanotubes }

\author{P.M.G.L. Ferreira, S. Kogikoski Jr., W. A. Alves, H. Martinho}

\email{herculano.martinho@ufabc.edu.br}

\affiliation{Centro de Ci\^{e}ncias Naturais e Humanas, Universidade Federal do ABC, Av. dos Estados 5001, Santo Andr\'{e}-SP, 09210-580, Brazil\\}

\begin{abstract}

The temperature dependence ($10-290$ K) of the low-frequency ($20-150$ cm$^{-1}$) Raman-active phonon modes of supercooled confined water in L,L-diphenylalanine micro/nanotubes was analysed. The isolated dynamics of a specific geometry of water cluster (pentamer) in supercooled confined regime was studied in detail. A particular mode concerning water-nanotube interaction was also probed. A fragile-to-strong transition at $204$ K was observed and related to the crossing of the Widom line. The critical exponent analyses of the relaxation rate data based on mode-coupling theory indicated perfect agreement among experimental data and theory. Our results are consistent with the existence of a second critical point of water.

\end{abstract}

\maketitle

The distinction between gas and liquid disappears above its critical point. At pressure and temperature above this point, the system is said to be in a fluid state (supercritical fluid)\cite{mcmillan2010fluid}. Supercritical fluids are recognized as possessing unique solvation properties that make them important technological materials\cite{mcmillan2010fluid}. Of particular interest is the behavior of water in confined spaces since it plays a key role in protein hydration since nanoscale fluctuations associated with the so-called Widom line can influence biological processes\cite{chu2009proteins,frenkel2002soft}.

Poole \textit{et al.} \cite{entropy5} presented a thermodynamically consistent molecular dynamical simulation study view regarding the global phase behavior of supercooled water. According to these authors, in the supercooled region just below the line of homogeneous ice nucleation, a critical point of liquid-liquid coexistence (LLCP) could exist that would eliminate the first-order transition line between low-density liquid (LDL) and high-density liquid (HDL) aqueous phases. Thus, liquid-liquid phase separation and the existence of the LLCP in water remains as a plausible hypothesis and requires further verification \cite{entropy}. The Widom line temperature $T_{W}$ corresponds to the loci of maxima of thermodynamic response function in the one-phase region beyond the LLCP proposed to exist in supercooled liquid water\cite{entropy5}.

Molecular dynamics simulations of the TIP$4$P/$2005$ model of water performed by Kumar \textit{et al.}\cite{kumar2013boson} indicated that the onset of the Boson peak in supercooled bulk water coincides with the crossover to a predominantly LDL-like below $T_{W}$. Gallo, Corradini and Roveri\cite{gallo2013fragile} studied the dynamical properties of aqueous solution of NaCl upon supercooling by molecular simulations. They found a crossover from a fragile (super-Arrhenius) to a strong (Arrhenius) behavior upon crossing the $T_{W}$ by both ionic solution and bulk water.

Experiments in the supercooling region are extremely difficult due to crystal nucleation processes. Thus experimental pieces of evidence concerning the different hypotheses supporting the existence of LLCP are hard to test\cite{MCTTIP4P}.

In confinement water can be more easily supercooled and studied in region of phase space where crystallization of bulk water cannot be avoided. Confined water in nanoporous silica have been extensively studied\cite{faraone2004fragile,LW,mallamace2007evidence,liu2005pressure,faraone2009single, bertrand2013deeply}. Faraone \textit{et al.}\cite{faraone2004fragile} confined water in synthesized nanoporous silica matrices MCM-41-S (pore diameters of $18$ and $14$ \AA) and interpreted the abrupt change of the relaxation time behavior observed by quasielastic neutron scattering at $T\sim 225$ K as the predicted fragile-to-strong liquid-liquid transition. Similar findings were reported by others (see, e.g.,\cite{LW,mallamace2007evidence}). Liu \textit{et al.} \cite{liu2005pressure} studied water confinement in MCM-41-S as function of pressure. They found that the transition temperature decreases steadily with an increasing pressure, until it intersects the homogenous nucleation temperature line of bulk water at a pressure of $1600$ bar. Above this pressure, it was no longer possible to discern the characteristic feature of the fragile-to-strong transition and it was elaborated that this point could be the possible second critical point of water. A hydrophobically modified MCM-41-SA matrix was studied in the $300-200$ K interval by Faraone  \cite{faraone2009single}. No evidence of the dynamic crossover, from a non-Arrhenius to an Arrhenius behavior was found corroborating others findings (see, e.g., ref. \cite{PhysRevE.76.021505}) reporting that the dynamic crossover takes place at a much lower temperature in water confined in hydrophobic confining media. For water in double-wall carbon nanotubes the transition occurred at $T\sim 190$ K. The water's tetrahedral hydrogen-bond network rule in the low temperature dynamic properties of confined water has also been revealed. The hydration-level dependence of the single-particle dynamics of water confined in the ordered mesoporous silica MCM-41 measured by Bertrand \textit{et al.}\cite{bertrand2013deeply} indicated that the dynamic crossover observed at full hydration was absent at monolayer hydration. The monolayer dynamics were significantly slower than those of water in a fully hydrated pore at ambient temperatures. 

Pieces of evidence for dynamical crossover for hydration water in proteins and other biomolecules have also been shown (see, e.g., the review of Mallamace \textit{et al.}\cite{mallamace2012dynamical}). Two special dynamical transitions, apparently of universal character, were observed in biomolecules (see, e.g., refs.\cite{lima2014boson,fogarty2013biomolecular,lima2012anharmonic,doster2010dynamical}). The first transition occurred at $T_{D}\sim 180-220$ K and was observed in macromolecules with a hydration level $h>0.18$. Based on the quasi-elastic neutron scattering measurements for lysozyme, Chen \textit{et al.} \cite{chen2006observation} interpreted the transition at $T_{D}$ as a fragile-to-strong dynamical crossover, where structured water makes a transition from a HDL to a LDL based on possible existence of LLCP\cite{entropy5}. However, Doster \textit{et al.} \cite{doster2010dynamical} showed no evidence of such a fragile-to-strong characteristic at $T_{D}$ for in fully deuterated C-phycocyanin protein. Thus, the exact nature of these dynamical transitions and the particular roles of protein and water are elusive.

To obtain direct evidence regarding the existence of the liquid-liquid phase transition in deeply supercooled water, we present results concerning the water confined in nanotubes of L-diphenylalanine (FF). This system is suitable for confining water in a controlled way. It can self-assemble into stiff, as well as chemically and thermally stable assembles in aqueous solutions. X-ray analysis have show that FF monomers crystallize with hydrogen-bonded head-to-tail chains in the form of helices with six dipeptide molecules per turn and a channel core with a diameter of $\sim10$ \AA, that is filled with water molecules \cite{wang2011effects}. Spectroscopic methods are powerful tools for exploring structural and dynamical properties of water \cite{mallamace2007evidence}. Raman spectroscopy is of special interest due to its high sensitivity and molecular specificity. This technique was employed to isolate and probe the microscopic dynamics of water inside FF studying.

The FF were prepared by using the liquid phase strategy, described by Reches and Gazit \cite{reches2003casting}. The Raman measurements were performed by using a triple spectrometer (T64000, HORIBA Jobin-Yvon) with a thermoelectrically cooled CCD detector (Synapse, HORIBA Jobin-Yvon). The 532 nm line of an optically pumped semiconductor laser (Verdi G5, Coherent) was used as the excitation source. The laser power at the sample was maintained at less than 15 mW with a spot size of $\sim100$ $\mu$m. The samples were cooled in the cold finger of a 4 K ultralow-vibration He closed-cycle cryostat (CS-204SF-DMX-20-OM, Advanced Research System). Measurements were conducted in a near backscattering configuration across an interval of $10<T<300$ K.  Each spectrum was deconvoluted to Pseudo-Voigt lineshape using the fityk software\cite{fityk}. The full width half maximum ($\Gamma$) and maximum frequencies ($\omega$) were computed for the vibrational modes of interest at each temperature.

\begin{figure}[t!]
\includegraphics[width=8.0cm]{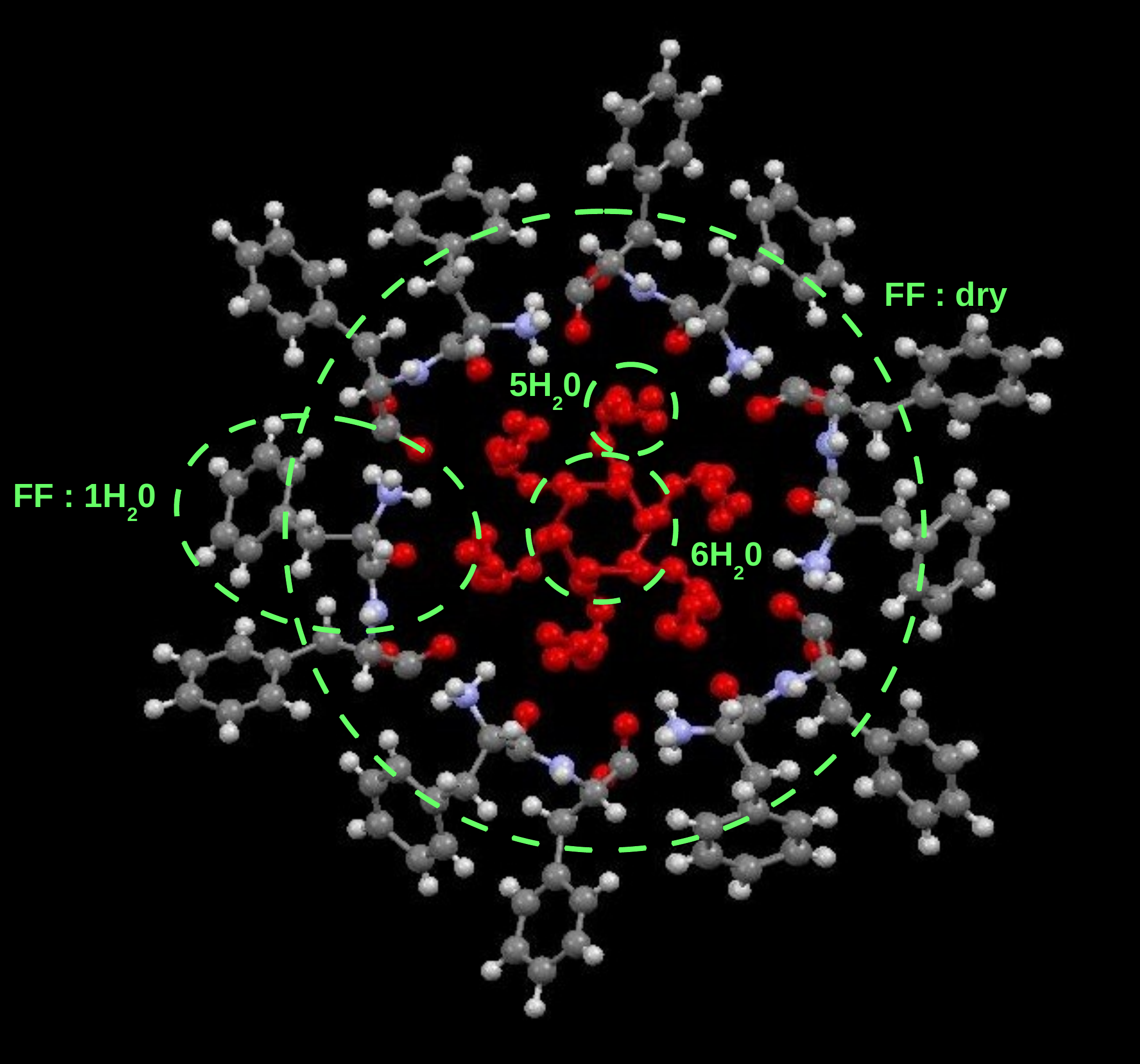}
\caption {(color online) Core motif of the formation of FF formed by six molecules of diphenylalanine arranged in a hexagonal pack, as seen from the $a$ and $b$ axes of the crystallographic unity cell.}
\label{fig:NtsClusters}
\end{figure}

The vibrational assignment was based on \textit{first-principle} vibrational calculations. Density Functional Theory (DFT)\cite{hohenberg1964inhomogeneous,kohn1965self} was used to obtain the equilibrium geometries and harmonic frequencies. Confinements were implemented in the Car-Parrinelo Molecular Dynamics program\cite{CPMD} by using the BLYP functional\cite{lee1988development} augmented with dispersion corrections for proper description of dispersion of the interactions\cite{von2005performance,lin2007library}. The cutoff energy of $75$ Ry was set in all simulations. The Raman-active vibrational modes calculations were performed on three subsystems of the FF hexagonal arrangement (Fig. \ref{fig:NtsClusters}) due to prohibitive computational cost to perform calculation on overall system. The atomic connectivity between the atoms in the simulation box was fixed in each subsystem:

\begin{itemize}
    \item[i)] The nanochannels without water that were obtained from the crystallographic data\cite{Ref12} (FF:dry);
    \item[ii)] The water clusters that were isolated from the nanochannel configuration (pentameter and hexamer arrangemeny);
    \item[iii)] The FF peptide interacting with one molecule of water (FF:1H$_{2}$O).
\end{itemize}
This approach enabled us to obtain directly the contributions of each isolated cluster to the Raman spectra. Due to the strong anharmonicity of the studied system, many vibrational modes corresponded to the calculated superposition of two or three eigenmodes.

\begin{figure}[tbh!]
\includegraphics[width=8.0cm]{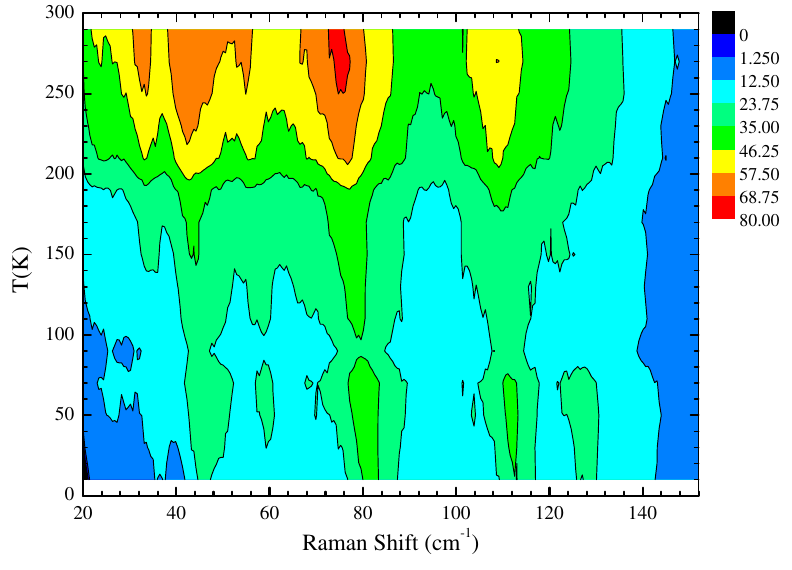}
\caption {(color online) Temperature dependence of the Raman spectra in the low frequency region.} \label{fig:resultado4}
\end{figure}

The temperature dependence ($10-290$ K) of the low wavenumber ($20-150$ cm$^{-1}$) Raman spectra of FF-MTs is shown on Fig.\ref{fig:resultado4}. The spectral variation above $T\sim 200$ K in the spectral window below $136$ cm$^{-1}$ is notewothy. Almost all phonons in this regions broaden and soften on heating. Also, modes at $36$; $45$; $50$; $60$; $82$; and $113$ experienced a noticeable intensity increasing. The detailed temperature dependence of the $82$ cm$^{-1}$ (water-nanotube mode) and $113$ cm$^{-1}$ (isolated heptamer water cluster mode) modes will be discussed on following.

\begin{figure}[tbh!]
\includegraphics[width=8.0cm]{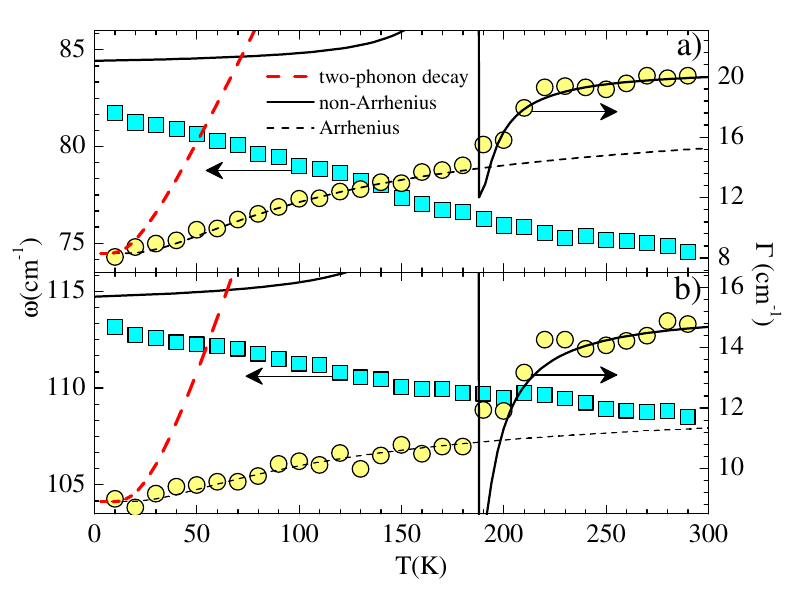}
\caption {(color online)Temperature dependency of $\omega$ (left scale) and $\Gamma$ (right scale) for vibration modes at $82$ cm$^{-1}$ (a) and $113cm^{-1}$ (b). The dasehd, solid, and short-dashed lines are fit to two-phonon decay, non-Arrhenius (eq. \ref{eqn:VTF}), and Arrenius (eq. \ref{eqn:Arrhenius}) expressions, respectively. The fitting parameters were indicated in the text. } \label{fig3}
\end{figure}

The left scale of Fig.\ref{fig3}a) and b) show $\omega(T)$ for $82$ and $113$ modes, respectively. Both presented a smooth softening on heating as result of crystal lattice thermal expansion  without signature of structural phase transition\cite{lima2012anharmonic}. Notwithstanding $\Gamma(T)$ displayed a clear anomaly close to $T\sim 200$ K for both modes (right scale of Fig.\ref{fig3} a) and b)). Increasing temperature $\Gamma(T)$ increases indicating shorter phonon relaxation time at high temperatures since $\Gamma$ is proportional to the relaxation rate $\Gamma \propto \tau^{-1}$.

The mechanisms by which phonon in solid can be scattered are anharmonic Umklapp process including boundary, phonon-impurity, electron-phonon, phonon-phonon scatterings\cite{dove1993introduction}. Each of these mechanism is associated with relaxation time which is inversely proportional to their relaxation rate. The total relaxation rate of phonon is given by Matthiessens rule: $\tau^{-1}= \sum_{i} \tau^{-1}_{i}$. Usually the temperature dependence of optical phonons is dominated by phonon-phonon anharmonic decay\cite{balkanski1983anharmonic}. The two-phonon decay process behavior for $\Gamma(T)$ expected for $82$ and $113$ cm$^{-1}$ phonons (following eq.$3.4$ of ref.\cite{balkanski1983anharmonic}) is shown on Fig.\ref{fig3}. One could conclude that these phonons were only weakly damped by anharmonic interactions since $\Gamma(T)<<\Gamma_{two-ph}$. Thus the two-phonon processes contributed only to a constant linewidth at $T\rightarrow 0$ limit. One possible reason is the absence of acoustic or optical phonons in the dispersion relation to match the conservation energy requirement. Thus, others possible decay channels could be explored.

The relaxation behavior of a deeply supercooled liquid is generally described by the viscosity-related main relaxation process ($\alpha$-kind) and one or several secondary relaxation processes ($\beta$-kind). For $\alpha$ processes  $\tau_{\alpha}$ usually displays some degree of non-Arrhenius (or fragile) temperature dependence\cite{grzybowska2010dynamic} being well-described by the Vogel-Fulcher-Tamman (VTF) law. The phonon relaxation rate in this case will be:

\begin{equation}\label{eqn:VTF}
\Gamma_{fragile}=\Gamma_{0}+ e^{-\frac{BT_{0}}{T-T_{0}}}
\end{equation}
where $\Gamma_{0}$ is a constant related to the residual two-phonon anharmonic decay, $B$ is a constant that provides a measure of fragility and $T_0$ is the ideal glass transition temperature. The strong (or Arrhenius) temperature dependence will be

\begin{equation}\label{eqn:Arrhenius}
\Gamma_{strong}=\Gamma_{0}+e^{-\frac{E_{A}}{k_{B}T}}
\end{equation}
where $E_A$ is the activation energy for the relaxation process and $k_B$  is the Boltzmann constant.

The fragile-to-strong transition is considered as a signature of the fluid crossing the Widom line (liquid-liquid transition). To investigate the structural relaxation mechanisms in FF $\Gamma(T)$ experimental data were fitted to eqs. \ref{eqn:VTF} and \ref{eqn:Arrhenius} as shown in Fig.\ref{fig3}. Below $T\sim 200$ K the data for both modes were well-described by the strong behavior becoming fragile above this crossover temperature. We notice that was not possible fit the overall data interval to one single expression. The existence of two relaxation regimes can be understood in terms of the so-called "cage effect". After the initial ballistic decay the system entered the $\beta$-relaxation region at an intermediate time scale and the $\alpha$-relaxation region at a long time scale. As temperature decreases, particles are trapped in the transient cage that is formed by their nearest neighbors. After the ballistic time region, the particles remain trapped in the cage. Finally, when the cage relaxes, the particles are free to diffuse again \cite{MCTTIP4P}. The best parameters found fitting data to eqs. \ref{eqn:VTF} and \ref{eqn:Arrhenius} were $E_{A}=0.214$ kcal/mol, $B=0.0478$, $T_{0}=187$ K. $\Gamma_{0}$ was fixed to $\Gamma_{0}= \lim_{T to 0} \Gamma(T)$. 

The crossover temperature $T_{W}$ was obtained from the expression\cite{mallamace2008transport}
\begin{equation}
\frac{1}{T_{W}}=\frac{1}{T_{0}}-\frac{B k_{B}}{E_{A}}
\end{equation}
and found to be $T_{W}= 204$K.

The dynamic anomaly of viscosity and the structural relaxation time in water have often been explained with mode-coupling theory (MCT) \cite{debenedetti1996metastable,PhysRevE.60.6757,0953-8984-15-45-L03}. MCT predicts that relaxation proceeds in essentially two steps in glass forming liquids at high temperatures ($T\gg T_{c}$) a fast $\beta$ relaxation step and a slow $\alpha$ relaxation step. The latter is the primary relaxation and correlates to the temperature variation of shear viscosity. According to MCT the fast relaxation have no temperature variation for $T>T_{c}$. The $\alpha$ relaxation maintains its spectral form and amplitude, but its characteristic time $\tau_{\alpha}$ follows the critical temperature behavior \cite{sokolov1995dynamics} $\tau_{\alpha} \propto (T/T_{c}-1)^{-\gamma}$  where $\gamma$ is the critical exponent. Thus, we could infer that the MCT prediction for the temperature dependence of linewidth follows the power law 

\begin{equation}\label{eqn:mct}
\Gamma(T)\sim(T/T_{c}-1)^{\gamma}
\end{equation}
$T_{c}$ is the critical temperature that marks the changes from a regime where relaxation process are mastered by breaking and reforming the cages to a regime where the cages are frozen and diffusion is attained through hopping \cite{MCT2}. It corresponds to the glassy transition temperature within this theory framework. We will test the previsions of MCT theory associating $T_{c}$ to $T_{W}$.

\begin{figure}[t!]
\includegraphics[width=8.0cm]{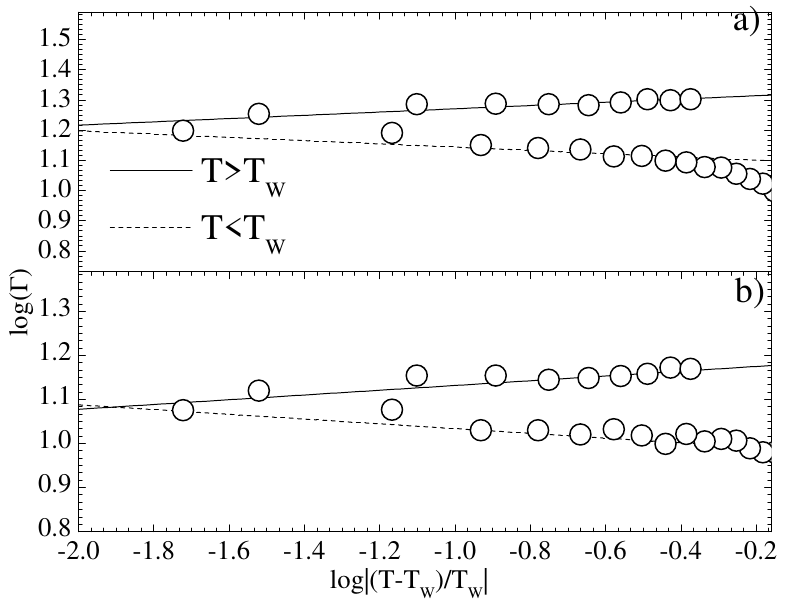}
\caption {$\Gamma$ as a function of reduced temperature with $T_{W} = 204$K in log-log scale for the vibration modes at (a) $82$ cm$^{-1}$ and (b) $113$ cm$^{-1}$.} \label{fig:MCT}
\end{figure}

Figure \ref{fig:MCT}a) and b) show a linearized log-log plot of $\Gamma$ as a function of reduced temperature for the bands $82$ cm$^{-1}$ and $113$ cm$^{-1}$, respectively. The experimental data below and above $T_{W}$ were separately fitted to a linear behavior. The agreement was perfect below $(T-T_{W})/T_{W}=-0.4$ for both data and furnished $\gamma=0.054$ for the critical exponent.This value is exactly that predicted by the self-consistent approximation of MCT \cite{bhattacharjee1981crossover, Ohta01051976}. It is important to notice that $\gamma=0.070$ whether one considers vertex corrections \cite{PhysRevA.14.884}. At $T>T_{W}$ the excellent agreement occurred in the overall range. 

We presented, to the best of our knowledge, the first reported on literature to study the isolated dynamics of a specific geometry of water cluster (pentamer) in supercooled confined regime. We showed that the supercooled confined water in FF exhibit a fragile-to-strong transition at $204$ K. From our analysis we concluded that this temperature corresponds to the Widom temperature $T_{W}$ which supports the theory of Poole \textit{et al.} \cite{entropy5} in which there is a liquid-liquid transition to supercooled water and that it ends at a critical point. The divergence behavior of $\Gamma(T)$ at $T_{W}$ furnished a critical exponent of $\gamma = 0.54$ in perfect agreement with that predicted by MCT. Our results are consistent with the existence of a second critical point of water in this system.

\textbf{Acknowledgements} The authors are grateful to the Brazilian agencies FAPESP, CAPES, and CNPq  for financial support and to the Multiuser Central Facilities of UFABC for experimental support.

\bibliography{mybib2}

\end{document}